\documentclass[11pt,a4paper]{article}
\pdfoutput=1

\usepackage{jheppub}
\usepackage{latexsym}
\usepackage{multirow}
\usepackage{color}
\usepackage[usenames,dvipsnames,svgnames,table]{xcolor}

\usepackage{graphicx}% Include figure files
\usepackage{epsfig}  % Include figure files
\usepackage{epsf}    % Include figure files
\usepackage{dcolumn}% Align table columns on decimal point
\usepackage{bm}% bold math
\usepackage{dcolumn}% Align table columns on decimal point
\usepackage{textcomp}% Align table columns on decimal point
\usepackage{float}
\usepackage{subfig}
\usepackage{hypcap}
\usepackage[]{hyperref}

\begin{document}
\author{Abhish Dev, }
\author{P. Ramadevi}
\author{ and S. Uma Sankar}
\title{Non-Zero $\theta_{13}$ and $\delta_{CP}$ in a Neutrino Mass Model with $A_4$ Symmetry}
\affiliation{Department of Physics, Indian Institute of Technology Bombay,\\
Mumbai 400 076, India}

\emailAdd{abhishdev92@gmail.com}
\emailAdd{ramadevi@phy.iitb.ac.in}
\emailAdd{uma@phy.iitb.ac.in}

\begin{abstract}{
 In this paper, we consider a neutrino mass model based on $A_4$ symmetry. The spontaneous symmetry breaking in this model is chosen to obtain tribimaximal mixing in the neutrino sector. We introduce $Z_2 \times Z_2$ invariant perturbations in this model which can give rise to acceptable values of  $\theta_{13}$ and $\delta_{CP}$. Perturbation in the charged lepton sector alone can lead to viable values of $\theta_{13}$, but cannot generate $\delta_{CP}$. Perturbation in the neutrino sector alone can lead to acceptable $\theta_{13}$ and maximal CP violation. By adjusting the magnitudes of perturbations in both sectors, it is possible to obtain any value of $\delta_{CP}$.  

}\end{abstract}

\maketitle
\section{Introduction}
The discovery of neutrino oscillations has triggered a lot of experimental and theoretical effort to understand the physics of lepton masses and mixing. Since flavor mixing occurs due to the mismatch between the mass and flavor eigenstates, neutrinos need to have small non-degenerate masses \cite{Maki:1962mu, Nakamura:2010zzi}. During the past two decades, many neutrino oscillation experiments have been performed and the values of oscillation parameters are determined to a very good precision \cite{Tortola:2012te,Fogli:2012ua,GonzalezGarcia:2012sz}. 

Neutrino oscillation probabilities depend only on the mass-squared differences and the mixing angles. Hence these parameters are determined in
the neutrino oscillation experiments. The experimental data has shown two large mixing angles and one small mixing angle. This pattern is different from the case of quark mixing where all angles are small and the mixing matrix is close to identity. The lepton mixing matrix, called Pontecorvo-Maki-Nakagawa-Sakata (PMNS) matrix, is  approximately equal to the
tribimaximal (TBM) ansatz proposed in ref.~\cite{Harrison:2002er}. In this ansatz, the mixing angles have values tan$^{2}\theta_{12} =\frac{1}{2}$, $\theta_{13}= 0^\circ$, and sin$^2\theta_{23}=\frac{1}{2}$.

The TBM form of the PMNS matrix is 
\begin{equation}\label{TBM}
U_{PMNS} \simeq \left( \begin{array}{rrr}
2/\sqrt{6} & 1/\sqrt{3} & 0 \\
-1/\sqrt{6} & 1/\sqrt{3} & -1/\sqrt{2} \\
-1/\sqrt{6} & 1/\sqrt{3} & 1/\sqrt{2}
\end{array} \right) \equiv U_\mathrm{TBM},
\end{equation}
where $|U_{e3}| = \sin \theta_{13} = 0$.

Many recent experiments \cite{Abe:2011fz,An:2012eh,Ahn:2012nd} have shown that the value of $\theta_{13}$ is not only non-zero but is relatively high \cite{An:2013zwz}. The values of other mixing angles also have small deviations from the TBM values. Since $\theta_{13}$ is non-zero, the possibility of a CP violating phase ($\delta_{CP}$) in the lepton mixing matrix must be considered seriously. The neutrino oscillation experiments have found two mass-squared differences with very different magnitudes. The smaller mass-squared difference, denoted $\Delta m^2_{21} = m^2_2 - m^2_1$, is positive and is of the order of $10^{-5} \text{eV}^2$. The larger mass-squared difference, $\Delta m^2_{31} = m^2_3 - m^2_1$, is of the order of $10^{-3} \text{eV}^2$, but its sign is not known. This leads to two  possible mass hierarchies for neutrinos: normal hierarchy (NH) in which $\Delta m^2_{31}$ is positive and $m_1<m_2<m_3$ and inverted hierarchy (IH) where $\Delta m^2_{31}$ is negative and $m_3 < m_1<m_2$. Finding the sign of $\Delta m^2_{31}$ is a major goal in many experiments like INO~\cite{Blennow:2012gj, Ghosh:2012px}, ICECube-PINGU \cite{Ribordy:2013xea, Winter:2013ema}, and long baseline experiments \cite{Paley:2013sta, Agarwalla:2013txa}. The values of mixing angles and mass-squared differences from the global analysis of data is summarized in Table~\ref{tabvalues} \cite{Schwetz:2011zk}.

\begin{table}[htbp]
\begin{center}
\begin{tabular}{|c|c|}
\hline Parameter & mean$^{(+1 \sigma, +2 \sigma, +3 \sigma)}_{(-1 \sigma, -2 \sigma, -3 \sigma)}$ \\

\hline &\\
 $\Delta m_{21}^{2} [10^{-5}eV^{2}]$ & $7.60_{(-0.18,-0.34,-0.49)}^{(+0.19,+0.39,+0.58)}$ \\ 
\hline &\\
 $\Delta m_{31}^{2} [10^{-3}eV^{2}]$ & \text{\tiny{(NH)}}$2.48_{(-0.06,-0.12,-0.18)}^{(+0.05,+0.10,+0.16)}$,\\
&\\
&$\text{\tiny{(IH)}}-2.38_{(-0.06,-0.12,-0.18)}^{(+0.05,+0.10,+0.16)}$ \\ 
\hline &\\
 $\sin^2 \theta_{12}$ & $0.323_{(-0.016,-0.031,-0.045)}^{(+0.016,+0.034,+0.052)}$ \\ 
\hline &\\
 $\sin^2 \theta_{23}$ &\text{\tiny{(NH)}}$0.567_{(-0.128,-0.154,-0.175)}^{(+0.022,+0.047,+0.067)}$, \\
 &\\
 &\text{\tiny{(IH)}}$0.573_{(-0.043,-0.141,-0.170)}^{(+0.025,+0.048,+0.067)}$ \\ 
\hline &\\ $\sin^2 \theta_{13}$ &\text{\tiny{(NH)}} $0.0234_{(-0.002,-0.0039,-0.0057)}^{(+0.002,+0.004,+0.006)}$,\\
&\\
&\text{\tiny{(IH)}}$0.0240_{(-0.0019,-0.0038,-0.0057)}^{(+0.0019,+0.0038,+0.0057)}$ \\ 
\hline 
\end{tabular}
\caption{The values of mass-squared differences and mixing angles from the global fits \cite{Forero:2014bxa}. The numbers in the parenthesis are upper/lower uncertainties at (1$\sigma$, 2$\sigma$, 3$\sigma$) confidence level.} 
\label{tabvalues}
\end{center}
\end{table}

To accommodate the small masses of neutrinos in comparison to charged leptons and quarks, a novel mechanism involving Majorana nature of neutrinos, called seesaw mechanism, was introduced in \cite{Minkowski:1977sc, Yanag:1979, YGell:1979c, Mohapatra:1979ia}. In this mechanism, the right handed partners of neutrinos are introduced with Majorana masses at high scale. The neutrinos, in addition, have Dirac masses of the order of charged lepton masses. %The Dirac masses arise from the coupling of ordinary left handed neutrinos to the right handed partners via the Standard Model Higgs field.
The most general neutrino mass matrix is a $6 \times 6$ matrix in the space of three left-handed and three right-handed neutrino fields. A diagonalization of this matrix leads to the generation of small Majorana masses for left-handed neutrinos. A common approach to obtain the observed mixing pattern is to constrain the structure of interaction Lagrangian, which gives rise to the mass matrix, using a discrete non-abelian flavor symmetry~\cite{Mohapatra:1998ka,Wetterich:1998vh,King:2001uz,Grimus:2001ex,Ohlsson:2002na,Ohlsson:2002rb,
Babu:2002dz,Kitabayashi:2002jd,Grimus:2003kq}. 
Many such models are constructed using discrete, non-abelian groups like $A_4$ 
\cite{Babu:2002dz,Ma:2001dn,Altarelli:2005yp,Altarelli:2005yx,He:2006dk} and $S_4$ 
\cite{Mohapatra:2003tw,Hagedorn:2006ug,Ma:2005pd}. In particular, it was shown in  
\cite{Altarelli:2005yp, Altarelli:2005yx, He:2006dk} that models based on $A_4$ symmetry can lead to the prediction of 
tribimaximal mixing. Being the smallest group with an irreducible triplet representation, $A_4$ has been popular group for 
neutrino mass models since its introduction in  ref. \cite{Ma:2001dn}.

 In the wake of $\theta_{13}$ measurement, it is necessary to modify the models predicting TBM pattern \cite{Acosta:2012qf, Acosta:2014dqa, Sierra:2013ypa, Sierra:2014hea}. Two major approaches to incorporate the necessary modifications are vacuum misalignment and symmetry breaking via perturbation terms. All models based on discrete symmetry groups require a special vacuum alignment condition to obtain tribimaximal mixing. A deviation from this, that is
 , a vacuum misalignment can lead to deviations from TBM pattern~\cite{Barry:2010zk, Grossman:2014oqa}. 
 Another way to generate deviations from TBM pattern is to add symmetry breaking terms which break the symmetry completely or partially \cite{He:2006dk, Hernandez:2013dta}. It is common to have  different residual symmetries in charged lepton and neutrino 
 sectors after such a perturbation.

 In this paper, we will consider modifications of a model based on $A_4$ group proposed in \cite{He:2006dk}. TBM pattern is obtained in this model by breaking $A_4$ symmetry spontaneously to $Z_3$ in the charged lepton sector and to $Z_2$ in the neutrino sector. We first introduce a $Z_2\times Z_2$ invariant complex  perturbation in the charged lepton sector only. This perturbation leads to non-zero value for $\theta_{13}$, small deviations in the values of $\theta_{12}$ and $\theta_{23}$, but does not lead to any CP violation. If a real $Z_2 \times Z_2$ perturbation is introduced in the neutrino sector only, viable values of $\theta_{13}$ and maximal CP violation are obtained. By introducing perturbations in both the charged lepton and the neutrino sectors, it is possible to obtain any value of $\delta_{CP}$ by adjusting their relative strengths.

\section{The $A_4$ Model}

$A_4$ is the group of even permutations on four elements and is the smallest group with a three dimensional irreducible representation which makes it a popular group in neutrino mass modelling. This group has three  1-dimensional irreducible representations and one  3-dimensional irreducible representation.% $A_4$ can be written as a semi-direct product of its normal subgroup $Z_2\times Z_2$ and $Z_3$. This allows us to write any element of $A_4$ as a product of an element from $Z_2\times Z_2$ and $Z_3$.
 There are two popular approaches to study the three dimensional irreducible representation: the Ma-Rajasekaran (M-R) approach~\cite{Ma:2001dn} which makes all the $Z_2\times Z_2$ elements diagonal and the Altarelli-Feruglio (A-F) approach~\cite{Altarelli:2005yx} in which the $Z_3$ elements are diagonal. We will use the M-R convention in our discussion. $A_4$ has four classes denoted by C1, C2, C3, and C4. The $3 \times 3$ matrix representations of the $A_4$ elements in each of these classes are:
\begin{eqnarray}
\text{C1}:&\left(
\begin{array}{ccc}
 1 & 0 & 0 \\
 0 & 1 & 0 \\
 0 & 0 & 1
\end{array}
\right),\nonumber\\
\text{C2}:&\left(
\begin{array}{ccc}
 1 & 0 & 0 \\
 0 & -1 & 0 \\
 0 & 0 & -1
\end{array}
\right) , \left(
\begin{array}{ccc}
 -1 & 0 & 0 \\
 0 & 1 & 0 \\
 0 & 0 & -1
\end{array}
\right), \left(
\begin{array}{ccc}
 -1 & 0 & 0 \\
 0 & -1 & 0 \\
 0 & 0 & 1
\end{array}
\right),\\
\text{C3}:&\left(
\begin{array}{ccc}
 0 & 1 & 0 \\
 0 & 0 & 1 \\
 1 & 0 & 0
\end{array}
\right), \left(
\begin{array}{ccc}
 0 & -1 & 0 \\
 0 & 0 & -1 \\
 1 & 0 & 0
\end{array}
\right), \left(
\begin{array}{ccc}
 0 & -1 & 0 \\
 0 & 0 & 1 \\
 -1 & 0 & 0
\end{array}
\right),\left(
\begin{array}{ccc}
  0 & 1 & 0 \\
 0 & 0 & -1 \\
 -1 & 0 & 0
\end{array}
\right),\nonumber\\
  \text{C4}:&\left(
\begin{array}{ccc}
 0 & 0 & 1 \\
 1 & 0 & 0 \\
 0 & 1 & 0
\end{array}
\right),\left(
\begin{array}{ccc}
 0 & 0 & -1 \\
 1 & 0 & 0 \\
 0 & -1 & 0
\end{array}
\right),\left(
\begin{array}{ccc}
 0 & 0 & 1 \\
 -1 & 0 & 0 \\
 0 & -1 & 0
\end{array}
\right)\left(
\begin{array}{ccc}
 0 & 0 & -1 \\
 -1 & 0 & 0 \\
 0 & 1 & 0
\end{array}\nonumber
\right).
\end{eqnarray}

The $Z_3$ elements in this group are
\begin{equation}
\left(
\begin{array}{ccc}
 1 & 0 & 0 \\
 0 & 1 & 0 \\
 0 & 0 & 1
\end{array}
\right),\left(
\begin{array}{ccc}
 0 & 1 & 0 \\
 0 & 0 & 1 \\
 1 & 0 & 0
\end{array}
\right),\left(
\begin{array}{ccc}
 0 & 0 & 1 \\
 1 & 0 & 0 \\
 0 & 1 & 0
\end{array}
\right).
\end{equation}

The $Z_2\times Z_2$ elements in this group are

\begin{equation}\label{z2gen}
\left(
\begin{array}{ccc}
 1 & 0 & 0 \\
 0 & 1 & 0 \\
 0 & 0 & 1
\end{array}
\right),\left(
\begin{array}{ccc}
 1 & 0 & 0 \\
 0 & -1 & 0 \\
 0 & 0 & -1
\end{array}
\right) , \left(
\begin{array}{ccc}
 -1 & 0 & 0 \\
 0 & 1 & 0 \\
 0 & 0 & -1
\end{array}
\right), \left(
\begin{array}{ccc}
 -1 & 0 & 0 \\
 0 & -1 & 0 \\
 0 & 0 & 1
\end{array}
\right).
\end{equation}
In this section, we will discuss the details of a type-I seesaw model based on $A_4$ group proposed in ref.\cite{He:2006dk}. We limit ourselves to the leptonic sector of the model. The fields in this sector are the three left-handed $SU(2)$ gauge doublets, three right-handed charged-lepton gauge singlets, and three right-handed neutrino gauge singlets. They are assigned to various irreducible representations of the $A_4$ group. In addition, there are four Higgs doublets, $\phi_i$ (i = 1,2,3) and $\phi_0$, and three scalar singlets $\chi_i$ (i=1,2,3). The assignments of the fields under various groups, are given in Table~\ref{assign}.
\begin{table}
    \begin{tabular}{ | l | l | l | p{5cm} | l |}
    \hline
     & $ SU(2) $ & $  U(1) $ & $ A_4 $ & 
   \\ \hline
    $ D_{iL} $ & $ \underline {\frac{1}{2}} $ & Y=-1 & $ \underline{3}  $ & left-handed doublets 
   \\ \hline
    $ l_{iR} $ & $ \underline {0} $ &  Y=-2 & $ \underline{1}\oplus \underline{1}'\oplus \underline{1}'' $ & right-handed charged lepton singlets
   \\ \hline
    $ \nu_{iR} $ & $ \underline {0} $ &  Y= 0 & $ \underline{3} $ & right-handed neutrino singlets 
   \\ \hline
    $ \phi_i $ & $ \underline {\frac{1}{2}} $ &  Y= 1  & $ \underline{3} $ & Higgs doublet
   \\ \hline
    $ \phi_0 $ &  $ \underline {\frac{1}{2}} $ &  Y= 1  & $ \underline{1} $ &  Higgs doublet
    \\ \hline
    $ \chi_i $ & $ \underline {0} $ &  Y= 0  & $ \underline{3} $ &  real gauge singlet
    \\ \hline       
    \end{tabular}
\caption{Assignments of lepton and scalar fields to various irreps of $ SU(2) $,  $U(1)$, and $A_4$.}
\label{assign}
\end{table}

By using the Clebsh-Gordon decomposition of $A_4$ tensor products, the complete $G_{SM}\otimes A_4$ invariant ($G_{SM}$ is the standard model gauge symmetry) Yukawa Lagrangian for the leptonic sector can be written as~\cite{Grimus:2011mp}
\begin{equation}
 \mathcal{L}_{\text{Yukawa}}= \mathcal{L}_{\text{CL Dirac}} + \mathcal{L}_{\text{N Dirac}} + \mathcal{L}_{\text{N Majorana}}.
 \end{equation} 
 The individual terms of this equation are given by 
 \begin{eqnarray} 
 \mathcal{L}_{\text{CL Dirac}}=&&-\left[h_1(\bar{D}_{1L}\phi_1+\bar{D}_{2L}\phi_2+\bar{D}_{3L}\phi_3)l_{1R} \right. \nonumber\\  
  &&+h_2(\bar{D}_{1L}\phi_1+\omega^2\bar{D}_{2L}\phi_2+\omega\bar{D}_{3L}\phi_3)l_{2R}\\ 
   &&\left. +h_3(\bar{D}_{1L}\phi_1+\omega\bar{D}_{2L}\phi_2+\omega^2\bar{D}_{3L}\phi_3)l_{3R}\right] +  \text{h.c.}, \nonumber
\end{eqnarray}
where $\omega$ is the cube root of unity,
\begin{equation}
\mathcal{L}_{\text{N Dirac}}=-h_0(\bar{D}_{1L}\nu_{1R}+\bar{D}_{2L}\nu_{2R}+\bar{D}_{3L}\nu_{3R})\tilde{\phi_0}
+\text{h.c.},
\end{equation}
and
     \begin{eqnarray} 
\mathcal{L}_{\text{N Majorana}}=&&-\frac{1}{2} \left[M(\nu_{1R}^TC^{-1}\nu_{1R}+\nu_{2R}^TC^{-1}\nu_{2R}+\nu_{3R}^TC^{-1}\nu_{3R})\right]+\text{h.c.}]\nonumber\\   
                                 &&- \frac{1}{2} \left[h_\chi(\chi_1(\nu_{2R}^TC^{-1}\nu_{3R}+\nu_{3R}^TC^{-1}\nu_{2R}) \right. \\
                                 &&+\chi_2(\nu_{3R}^TC^{-1}\nu_{1R}+\nu_{1R}^TC^{-1}\nu_{3R})\nonumber\\
                                 &&\left. + \chi_3(\nu_{1R}^TC^{-1}\nu_{2R}+\nu_{2R}^TC^{-1}\nu_{1R})\right],\nonumber                          
                                 \end{eqnarray}
where $C$ is the charge conjugation matrix. Here, $\mathcal{L}_{\text{CL Dirac}}$ contributes to the Dirac mass matrix
in the charged lepton sector,  $\mathcal{L}_{\text{N Dirac}}$ contributes to the Dirac mass matrix in the neutrino sector 
and $\mathcal{L}_{\text{N Majorana}}$ contributes to the Majorana mass matrix of the right handed neutrinos. 
$\mathcal{L}_{\text{Yukawa}}$ has an additional $U(1)_X$ symmetry \cite{He:2006dk}. Under this symmetry the fields
$D_{iL}$, $l_{iR}$ and $\phi_i$ have quantum numbers $X=1$ and all other fields have $X=0$. This symmetry forbids
the Yukawa terms of the form $\bar{D}_{L} \nu_R \tilde{\phi}_i$. These terms are invariant under $G_{SM} \times
A_4$ and contribute to the Dirac mass matrix of the neutrinos. Without the contribution of these terms, this
matrix retains the simple form needed to obtain the tribimaximal mixing.

Spontaneous symmetry breaking leads to the following scalar VEVs: $v_i$ for $\phi_i$, $w_i$ for $\chi_i$, and $v_0$ for $\phi_0$. With these VEVs, we obtain the different mass terms to be
\begin{equation}
-\bar{l}_{L}M_l^0l_{R} - \bar{\nu}_{L}M_{D}\nu_{R} + \frac{1}{2}\nu_R^TC^{-1}M_{R}\nu_{R} + h.c. , 
\end{equation}
where
\begin{equation}\label{lepmass}
M_l^0 =\text{  }\left(\text{  }
\begin{array}{ccc}
 h_1v_1 & h_2 v_1 & h_3v_1 \\
 h_1 v_2 & h_2v_2\omega ^2 & h_3v_2\omega  \\
 h_1 v_3 & h_2 v_3\omega  & h_3 v_3\omega ^2
\end{array}
\right),\text{        }M_R\text{  }=\text{  }\left(\text{  }
\begin{array}{ccc}
 M & h_{\chi }w_3 & h_{\chi }w_2 \\
 h_{\chi }w_3 & M & h_{\chi }w_1 \\
 h_{\chi }w_2 & h_{\chi }w_1 & M
\end{array}
\right),
\end{equation}
and $M_D$ = $h_0 v_0 I$. Tribimaximal mixing requires a special vacuum alignment given by
\begin{equation}
v_1 = v_2 = v_3 = v,\text{ } w_1 = w_3 = 0,\text{and } h_\chi w_2 = M^{'}.
\end{equation}
The charged lepton mass matrix $M_l^0$ can be put in a diagonal form by the transformation 
\begin{equation}
U_\omega M_l^0 I = \left(\text{  }
\begin{array}{ccc}
 \sqrt{3}v h_1 & 0 & 0 \\
 0 & \sqrt{3}v h_2 & 0 \\
 0 & 0 & \sqrt{3}v h_3
\end{array}
\right)  \hspace{0.2in}\text{where } U_\omega = \frac{1}{\sqrt{3}}\left(\text{  }
\begin{array}{ccc}
 1 & 1 & 1 \\
 1 & \omega  &  \omega ^2 \\
 1 & $ \text{} $ \omega ^2 & \omega 
\end{array}
\right).
\end{equation}
The Majorana mass matrix $M_R$ is diagonalized by an orthogonal transformation 
\begin{equation}
U_{\nu }M_RU_{\nu }^{\dagger}=\left(
\begin{array}{ccc}
 M + M' & 0 & \text{  }0 \\
 \text{ }0 & M & \text{  }0 \\
 \text{ }0 & 0 & M - M'
\end{array}
\right) 
~~~~~~~~~ \text{ where }
 U_\nu = \left(
\begin{array}{ccc}
 \frac{1}{\sqrt{2}} & 0 & -\frac{1}{\sqrt{2}} \\
 0 & 1 & $ \text{} $ 0 \\
 \frac{1}{\sqrt{2}} & 0 & $ \text{} $ \frac{1}{\sqrt{2}}
\end{array}
\right).
\end{equation}
The PMNS matrix is now obtained to be tribimaximal up to phases on both sides.
\begin{equation}
U = U_{\omega } U_{\nu  }=\text{   }\left(\text{  }
\begin{array}{ccc}
 1 & 0 & 0 \\
 0 & \omega  & 0 \\
 0 & 0 & $ \text{} $ \omega^{2} 
\end{array}
\right)\left(
\begin{array}{ccc}
 \frac{ 2}{\sqrt{6}} & \frac{1}{\sqrt{3}} & $ \text{} $ 0 \\
 \frac{-1}{\sqrt{6}} & \frac{1}{\sqrt{3}} & -\frac{1}{\sqrt{2}} \\
 \frac{-1}{\sqrt{6}} & \frac{1}{\sqrt{3}} & $ \text{} $ \frac{1}{\sqrt{2}}
\end{array}
\right)\left(
\begin{array}{ccc}
 1 & 0 & $ \text{} $  0 \\
 0 & 1 & $ \text{} $  0 \\
 0 & 0 & -i
\end{array}
\right).
\end{equation}
 The vacuum alignment for scalar fields spontaneously breaks the $A_4$ symmetry in the charged lepton sector (coupling only with $\phi_i$) to $Z_3$ subgroup. In the neutrino sector (coupling with $\chi$ and $\phi_0$), the residual symmetry is $Z_2$. The Lagrangian lacks a common symmetry as there is no subgroup between $Z_2$ and $Z_3$. A novel feature of the model is that the diagonalizing matrix is completely determined by the symmetry, but the lepton masses are given by the arbitrary coupling constants $h_i$ ($i=0,1,2,3$). The seesaw mechanism generates small masses for the left handed neutrinos given by $M_D^TM_R^{-1}M_D$. The masses of the left-handed neutrinos then become $m_D^2/(M+M')$,$m_D^2/M$, and $m_D^2/(M-M')$,
satisfying the relation $2 m_2^{-1} = m_3^{-1} + m_1^{-1}$ \cite{Barry:2010yk}. 
 For $M^{\prime}\ll M$, a quasi degenerate spectrum is obtained.
\section{Perturbation in Charged Lepton Sector}
In the model discussed till now, the PMNS matrix has the tribimaximal form with zero $\theta_{13}$ and no CP violation.
To generate non-zero values for these, we add small perturbations to the above model. We first introduce a symmetry breaking
term in the charged lepton sector which is invariant under the subgroup $Z_2\times Z_2$.  %However, since the $A_4$ symmetry
%is spontaneously broken to $Z_3$ in charged lepton sector and $Z_2$ in neutrino sector,
%the perturbed lagrangian will have no symmetry in the charged lepton sector but will preserve the
%$Z_2$ symmetry in the neutrino sector.
In order to construct such a
perturbation, it is required to know the breaking pattern of $A_4$ irreducible representations  into $Z_2\times Z_2$
irreducible representations. The group $Z_2 \times Z_2$ is the normal subgroup of $A_4$ with four elements. It has one
trivial singlet representation $\mathbf {\underline {\hat 1}} ( 1, 1, 1, 1)$ and three non-trivial singlet representations,
viz. $\mathbf {\underline {\hat 1}'''} (1, 1, -1, -1)$, $\mathbf {\underline {\hat 1}''} (1, -1, 1, -1)$
and $\mathbf {\underline {\hat 1}'}(1, -1, -1, 1)$. The breaking of $A_4$ triplet into $Z_2 \times Z_2$ irreducible
representations can be readout from the diagonal matrix elements of $Z_2 \times Z_2$ in the M-R basis, shown in eq.(\ref{z2gen}). This is given as 
   \begin{eqnarray}
(\mathbf{\underline3})\text{ of }A_4 &\xrightarrow{\text{breaks into}}& (\mathbf {\underline {\hat 1}'''}\oplus \mathbf {\underline {\hat 1}''}\oplus\mathbf {\underline {\hat 1}'}) \text{ of } Z_2\times Z_2 \nonumber\\
   (\mathbf{\underline1}, \mathbf{\underline1'},\mathbf{\underline1''}) \text{ of }A_4 &\xrightarrow{\text{breaks into}}& \text{   }\hspace{0.35in}(\mathbf{\underline{\hat 1}})\hspace{0.45in}\text{of } Z_2\times Z_2.
    \end{eqnarray}   
The general $Z_2\times Z_2$ invariant perturbation can be written as
\begin{equation}\label{CLpert}
h_1\underset{(\underbar{\textbf{3}})}{\mathrm{\bar{D}_L}} M_1 \underset{(\underbar{\textbf{3}})}{\mathrm{\phi}}l\underset{(\hat{\underbar{\textbf{1}}})}{\mathrm{_{1R}}} 
+ h_2\underset{(\underbar{\textbf{3}})}{\mathrm{\bar{D}_L}} M_2 \underset{(\underbar{\textbf{3}})}{\mathrm{\phi}}l\underset{(\hat{\underbar{\textbf{1}}})}{\mathrm{_{2R}}} +h_3\underset{(\underbar{\textbf{3}})}{\mathrm{\bar{D}_L}} M_3 \underset{(\underbar{\textbf{3}})}{\mathrm{\phi}}l\underset{(\hat{\underbar{\textbf{1}}})}{\mathrm{_{3R}}} 
\end{equation}
where  $\bar{D}_{L}$, $\phi$ are the three-dimensional reducible representations 
of $Z_2\times Z_2$ and $l_R$'s are trivial singlets. For the perturbation to be $Z_2\times Z_2$ invariant
, the matrices $M_1, M_2$ and $M_3$ must commute with the matrices given in eq.~(\ref{z2gen}).
This is satisfied by any diagonal matrix.

%{\bf We see that the model is built in the basis where all $Z_2 \times Z_2$
%elements are diagonal which is important for the simplified choice of the perturbation matrix.
%The same model and perturbation can be written in any other basis obtained by a  unitary
%transformation of triplets while retaining the same mixing matrix, field assignments,
%and $Z_2 \times Z_2$ invariance. The structure of perturbation and charged lepton and
%neutrino mass matrices will be different in the new basis. The change of basis also modifies
%the Clesh-Gordon decomposition which is discussed in the appendix A.2.1 in ref. \cite{Barry:2010zk}.
%The choice of M-R basis reveals the $Z_2 \times Z_2$ invariance of the chosen perturbation in a simple way.}

It can be observed that introducing a multiplicative factor in the $i^{\text{th}}$ row of charged lepton mass matrix in eq.~(\ref{lepmass}) will introduce a reciprocal factor in the $i^{\text{th}}$ column of its diagonalizing matrix $U_\omega$. The $U_{e3}$  element of the PMNS matrix in the TBM form is zero because the 11 and 13 elements of $U_\omega$ are equal. 
The perturbation terms in eq.~(\ref{CLpert}) can disturb this balance and lead to non-zero $U_{e3}$. 
The value of $U_{e3}$ (and hence $\theta_{13}$) depends on the elements of $M_1$, $M_2$, and $M_3$. 
 In order to obtain a simple form for the perturbed charged lepton mass matrix $M_l$, we choose $M_i$s of 
the form $M_i = \text{diag} (\bar{z}, 0, \omega^{i-1}z)$ where $z$ is a complex number with $|z|\ll 1$. 
After spontaneous symmetry breaking, the resulting $M_l = M_l^0 + \Delta M_l$, where $M_l^0$ is given in eq. (\ref{lepmass}) and

\begin{equation}
\Delta M_l =\text{  }\left(\text{  }
\begin{array}{ccc}
 h_1v\bar{z} & h_2v\bar{z} & h_3v\bar{z} \\
0& 0 &0 \\
 h_1 vz & h_2 vz\omega  & h_3 vz\omega ^2
\end{array}
\right).
\end{equation}
Such a $\Delta M_l$ can arise from higher order effects of the theory. The form of $\Delta M_l$ is similar to the form 
of $\Delta M_{u,d}$ given in eq. (4.3) of \cite{He:2006dk}. In generating these higher order terms, the Higgs VEVs are 
unaffected. To simplify the phenomenological analysis, we parameterized 
all the six higher order terms in terms of a single number $z$. Note that there is no residual symmetry left in
the charged lepton sector after the spontaneous symmetry breaking. The perturbed matrix elements of $M_l$ introduce
reciprocal factors in the respective columns of $U_\omega$. Requiring $U_\omega$ to be unitary, we get $z$ to be 
 \begin{equation}\label{eq:18}
 z = -1 \pm \sqrt{1-s^2} + i s.
 \end{equation}
  We will retain the solution with $+$ sign in order to keep $|z|<1$. The perturbation strength is of the order $s$ which we take to be small.
 Using the parametrization $s$ =  $\sin \alpha$, we can transform $U_\omega$ to  
\begin{equation}
U_{\omega} = \frac{1}{\sqrt{3}}
\left(
\begin{array}{ccc}
 e^{ i \alpha} &1& \text{      }e^{ -i \alpha } \\
 e^{ i \alpha} &\omega& \text{            }\omega ^2 e^{- i \alpha } \\
 e^{ i \alpha } &\omega^2& \text{            }\omega  e^{- i \alpha }
\end{array}
\right).
\end{equation}
The PMNS matrix becomes
\begin{equation}\label{pmnsclpert}
U_{PMNS}=\frac{1}{\sqrt{3}}
\left(\text{  }
\begin{array}{ccc}
 1 & 1 & 1 \\
 1 & \omega & \omega^2 \\
 1 & \omega^2 & \omega
\end{array}
\right) \left(\text{  }
\begin{array}{ccc}
 e^{i \alpha} & 0 & 0 \\
 0 & 1 & 0 \\
 0 & 0 &e^{-i \alpha}
\end{array}
\right)
\left(
\begin{array}{ccc}
 \frac{1}{\sqrt{2}} & 0 & -\frac{1}{\sqrt{2}} \\
 0 & 1 & 0 \\
 \frac{1}{\sqrt{2}} & 0 & \frac{1}{\sqrt{2}}
\end{array}
\right).
\end{equation}
A similar structure for the PMNS matrix is discussed in refs.~\cite{BenTov:2012tg, Feruglio:2012cw, Chen:2013wba}. From the above equation, we can compute the perturbed values of the mixing angles to be 
\begin{alignat}{3}
&\sin^2\theta_{13}\text{ } &=&\text{    }\frac{2}{3}\sin^2\alpha &&= \frac{2s^2}{3}\notag\\
&\sin^2\theta_{12}\text{ } &=&\text{   }\frac{1}{2+\cos 2\alpha} &&=\frac{1}{3}+\frac{2 s^2}{9}+O(s^3)\\
&\sin^2\theta_{23}\text{ } &=&\text{ }\frac{2 + \cos 2\alpha + \sqrt{3}\sin 2\alpha}{2(2+\text{cos }2\alpha)}&&=\frac{1}{2}+\frac{s}{\sqrt{3}}+O(s^3).\notag
\end{alignat}
\begin{figure}[htbp]
\centering
\includegraphics[scale=0.8]{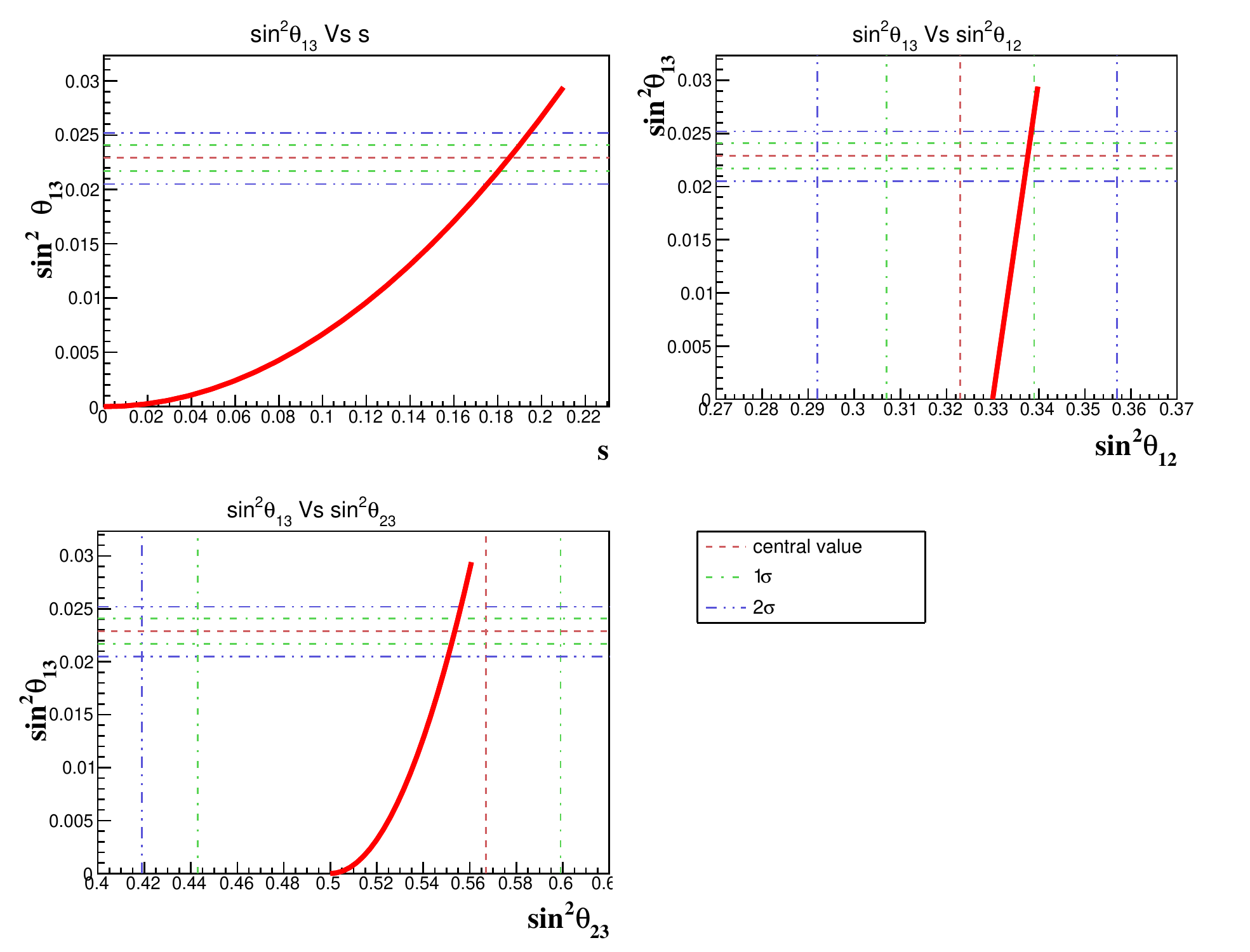}
\caption{ The plot of sine squared values of the mixing angles due to a $Z_2 \times Z_2 $ invariant perturbation in the charged lepton sector. Lines demarcating the central values and the 1$\sigma$ and 2$\sigma$ allowed regions are shown explicitly.}
\label{clpertangles}
\end{figure}
The sine squared values of mixing angles in this scheme are plotted in Figure~\ref{clpertangles}.

The perturbation parameter $s\sim 0.19$ leads to a very good fit for $\theta_{13}$. Such a value of $s$ also gives $\sin^2 \theta_{23}$ very close to the central value and $\sin^2\theta_{12}$ within $2\sigma$ range. Compared to their respective TBM values,  $\sin^2\theta_{12}$  changes very little ($\sim 5 \%$), whereas $\sin^2\theta_{23}$ receives an appreciable correction ($\sim 14 \%$).

We introduced perturbations in both the first and third rows of $M_l$. We chose these perturbations to be related to each other. This enabled us to keep the perturbation $s$ at the level of $10-20\%$. In principle, it is possible to choose the perturbing matrix $M_i$ = diag ($\bar{z}$, 0, 0). Such a perturbation modifies only the first row of $M_l$. Parametrizing $z$ in terms of $s$ as in eq.~(\ref{eq:18}), we can obtain the modified values of the mixing angles. With $s=\sin \alpha$, these values are
\begin{alignat}{3}
&\sin^2\theta_{13}\text{ } &=&\text{    }\frac{2}{3}\sin^2\frac{\alpha}{2} &&= \frac{s^2}{6}+O(s^4)\notag\\
&\sin^2\theta_{12}\text{ } &=&\text{   }\frac{1}{2+\cos\alpha} &&=\frac{1}{3}+\frac{s^2}{18}+O(s^4)\\
&\sin^2\theta_{23}\text{ } &=&\text{ }\frac{2 + \cos\text{ }\alpha + \sqrt{3}\sin \alpha}{2(2+\text{cos }\alpha)}&&=\frac{1}{2}+\frac{s}{2 \sqrt{3}}+O(s^3).\notag
\end{alignat}
In this case, the amount of perturbation should be double that of the previous case to obtain an acceptable value of $\theta_{13}$.

Given that we obtained viable values of $\theta_{13}$ we check if a CP violating phase $\delta_{CP}$ is also generated. However, we find that the Jarlskog invariant $J$ of the PMNS matrix in eq.~(\ref{pmnsclpert}) is zero. Hence, no CP violation can be generated by the perturbations considered here. So we look for other possible sources of CP violation and also non-zero $\theta_{13}$ in this model.  
\section{Perturbation in Neutrino Sector}
 In the previous section it was shown that a $Z_2 \times Z_2$ invariant perturbation in the charged lepton sector can give rise to viable $\theta_{13}$ but no CP violation. In this section, we add a similar perturbation in the neutrino sector and study its influence on $\theta_{13}$ and $\delta_{CP}$.  As in the case of the charged lepton sector, the perturbing matrix should be diagonal to satisfy the $Z_2 \times Z_2$ symmetry.  
%The value of $\delta_{CP}$ depends, in general, on the magnitudes of both the perturbations.
%By varying them appropriately, it is possible to obtain any value of $\delta_{CP}$. In particular, maximal CP violation requires zero perturbation in the charged lepton sector.  
We will derive expressions for $\theta_{13}$ and $\delta_{CP}$ as a function of the two perturbations and show that it is possible to obtain any value of $\delta_{CP}$. %In particular, maximal CP violation requires zero perturbation in the charged lepton sector.
It is shown that perturbation only in the neutrino sector leads to maximal CP violation.
 
\begin{figure}[htbp]
\centering
\includegraphics[scale=0.35]{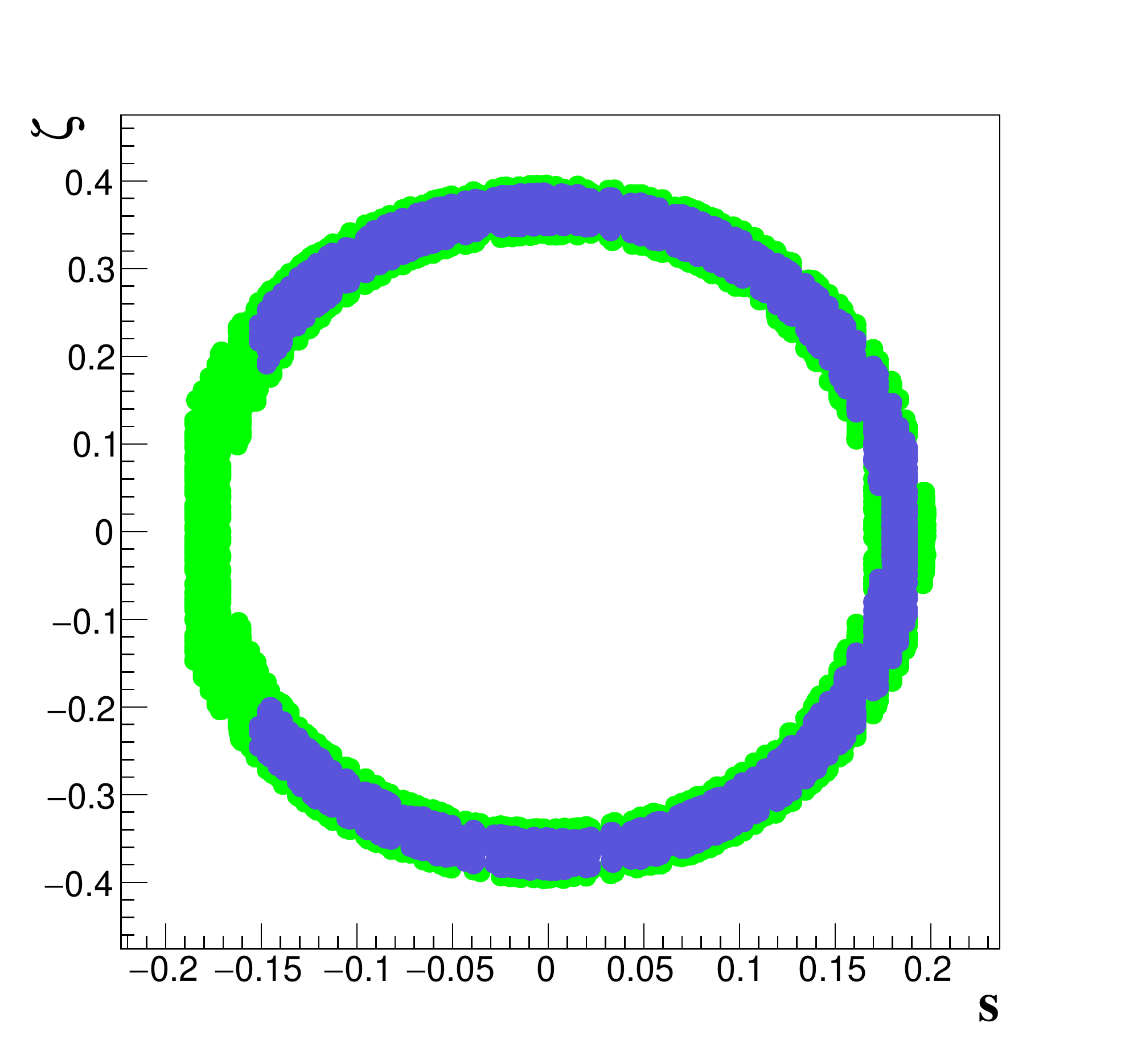}
\includegraphics[scale=0.5]{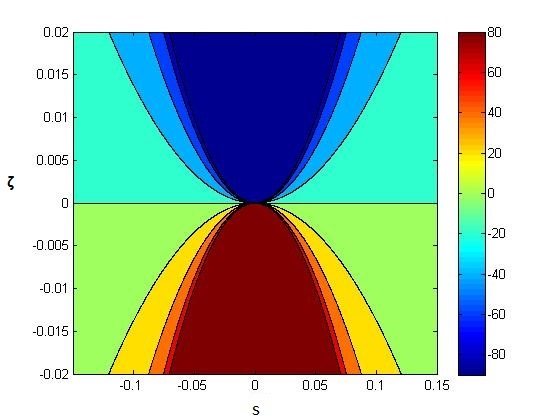}
\caption{Left panel: The points in $s-\zeta$ space which satisfy the 2$\sigma$ (blue band) and 3$\sigma$ (green band) constraints on $\sin^2 \theta_{12}$,
$\sin^2 \theta_{13}$ and $\sin^2 \theta_{23}$. Right panel: The value of $\delta_{CP}$ for different regions in the $s-\zeta$ space.}
\label{neutrCP}
\end{figure}

\begin{figure}[htbp]
\centering
\includegraphics[scale=0.38]{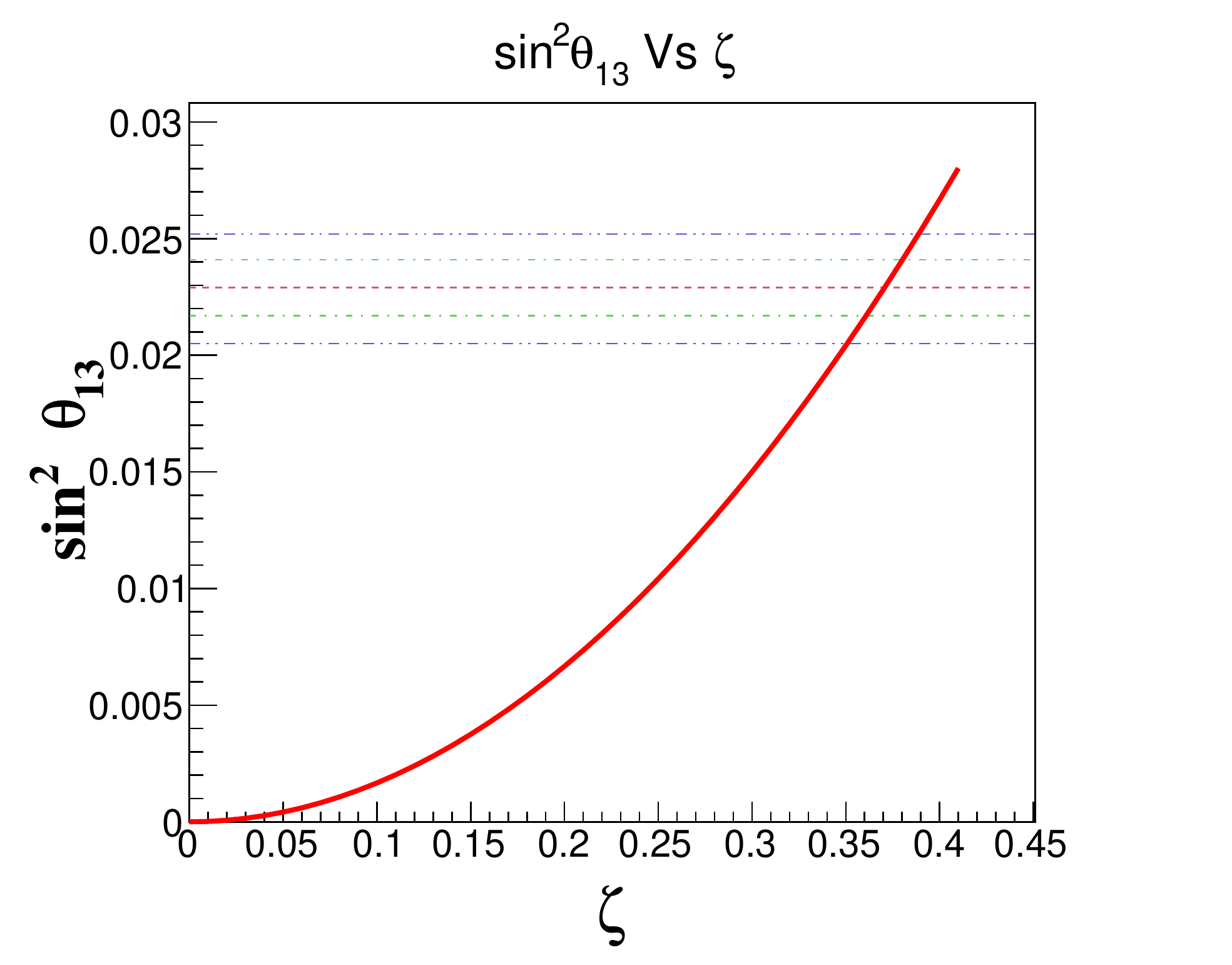}
\includegraphics[scale=0.35]{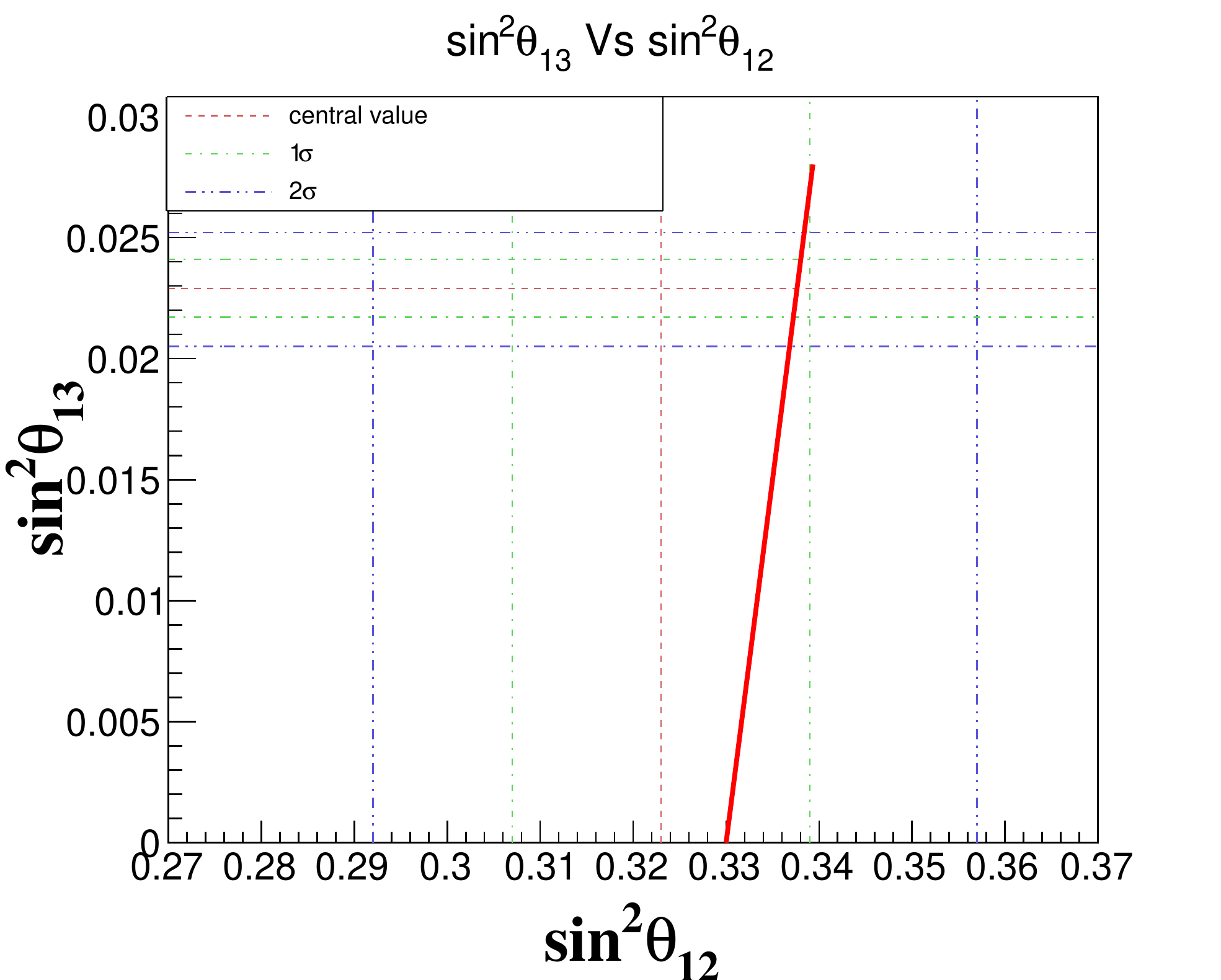}
\caption{The plot of sine squared vales of  mixing angles for maximal $\delta_{CP}$ through a $Z_2 \times Z_2 $ invariant perturbation with lines for 1$\sigma$ and 2$\sigma$ range.}
\label{neutrangles}
\end{figure}

 We observe that the diagonalizing matrix in the neutrino sector is a rotation matrix of angle $\pi/4$. A small imbalance in the degeneracy of $11$ and $33$ elements of $M_R$ in eq.~(\ref{lepmass}) shifts the rotation angle slightly away from $\pi/4$ \cite{He:2006dk}. Such an imbalance can be introduced by a $Z_2 \times Z_2$ invariant perturbation in the neutrino sector. We choose this perturbation to be \cite{He:2006dk, BenTov:2012tg, Feruglio:2012cw}
\begin{equation}
M\underset{(\underbar{\textbf{3}})}{\mathrm{\nu_R^T}} C^{-1}\left(\text{  }
\begin{array}{ccc}
  a & 0 & 0 \\
  0 & 0& 0 \\
  0 & 0 & - a
\end{array}
\right)\underset{(\underbar{\textbf{3}})}{\mathrm{\nu_R}}.
\label{4.1}
\end{equation} 
 The mass $M$ is a $A_4$ invariant soft term in the lagrangian. The perturbation in eq. (\ref{4.1}) can be introduced as an $A_4$ breaking
 but $Z_2 \times Z_2$ preserving soft term in the lagrangian. The perturbed Majorana mass matrix becomes 
\begin{equation}
\left(\text{  }
\begin{array}{ccc}
 M + a M & 0 & M' \\
 0 & M  & 0 \\
 M' & 0 & M - a M
\end{array}
\right).
\end{equation}
This matrix can be diagonalized by a rotation of angle $`x$', where $\tan 2x = M'/aM$. We will denote the perturbation in the neutrino sector by the dimensionless parameter $\zeta = aM/M' (\equiv \cot 2x)$.
The form of PMNS matrix after the combined perturbations in the charged lepton and the neutrino sectors is
\begin{equation}
U_{PMNS}=\frac{1}{\sqrt{3}}
\left(\text{  }
\begin{array}{ccc}
 1 & 1 & 1 \\
 1 & \omega & \omega^2 \\
 1 & \omega^2 & \omega
\end{array}
\right) \left(\text{  }
\begin{array}{ccc}
 e^{i \alpha} & 0 & 0 \\
 0 & 1 & 0 \\
 0 & 0 & e^{-i \alpha}
\end{array}
\right)\left(\text{  }
\begin{array}{ccc}
 \cos\text{ }x & 0 & -\sin\text{ }x \\
 0 & 1 & 0 \\
 \sin\text{ }x & 0 & \cos\text{ }x
\end{array}
\right).
\end{equation}
We recall that the perturbation in the charged lepton sector $s=\sin \alpha$. The Jarlskog invariant of this matrix can be found to be $\sqrt{3}\cos 2 x/18$ which vanishes for $x=\pi/4$. We obtain CP violation due to the  deviation of the  angle `$x$' from $\pi/4$ through the perturbation in the neutrino sector.
%The mass eigenvalues after the perturbation are,
%\begin{eqnarray}
%&m_1& = M + aM \cos 2x + M' \sin 2x,\nonumber \\
%&m_2&= M,\\
%&m_3&=  M - aM \cos 2x - M'\sin 2x.\nonumber
%\end{eqnarray}
%After making the substitution $Ma/M'$ = $\zeta $ ($=\cot2x$) and considering $\zeta < 1$, the modified %values of the mixing angles, to the order $\zeta^2$ and $s^2$, (where $s = \sin \alpha$ parameterizes the %perturbation in the charged lepton sector as discussed in the previous section) are
Expanding the expressions for the mixing angles up to order $\zeta^2$ and $s^2, $ we get
\begin{alignat}{3}
&\sin^2\theta_{13}\text{ } &=&\text{    }\frac{1}{3} (1-\cos 2 \alpha  \sin 2 x) &&= \frac{\zeta ^2}{6}+\frac{2}{3}s^2-\frac{\zeta ^2s^2}{3},\label{th13} \\
&\sin^2\theta_{12}\text{ } &=&\frac{1}{2+\cos 2 \alpha \sin 2 x}&&=\frac{1}{3}+\frac{\zeta ^2}{18}+\frac{2}{9}s^2-\frac{\zeta ^2s^2}{27},\\
&\sin^2\theta_{23}\text{ } &=&\text{   }\frac{2+\cos 2 \alpha  \sin 2 x+\sqrt{3} \sin 2 x \sin 2 \alpha }{4+2 \cos 2 \alpha  \sin 2 x} &&=\frac{1}{2}+\frac{s}{\sqrt{3}}-\frac{\zeta ^2s}{3 \sqrt{3}}.\label{th23}
\end{alignat}
From these values and the Jarlskog invariant, we obtain $\sin \delta_{CP}$ to be 
\begin{alignat}{2}
\sin\text{ }\delta_{CP}=\frac{\cos 2 x (2+\cos 2 \alpha \sin 2 x )}{\sqrt{\left(1-\cos^2 2 \alpha \sin^2 2 x\right) \left[4+4 \cos 2 \alpha \sin 2 x +(-1+2 \cos 4 \alpha) \sin^2 2 x\right]}}.\label{eqexact}
\end{alignat}
The expression in eq.~(\ref{eqexact}) is exact. We can obtain a simpler equation by expanding it in $\zeta$ and $s$ and keeping only the leading powers in the numerator and the denominator,
\begin{alignat}{2}
\sin\text{ }\delta_{CP} 
          &=-\frac{\zeta }{\sqrt{4 s^2+\zeta ^2-\frac{16 s^2 \zeta ^2}{3}}}.&\label{eqapprox}
\end{alignat}
The value of $\delta_{CP}$ goes to zero as $\zeta$ tends to zero, corresponding to no perturbation in the neutrino sector. For perturbation only in the neutrino sector, we have $s=0$ and $\delta_{CP} = \pm\pi/2$, depending on the sign of $\zeta$. % It is interesting to note that this value is the best fit point for the $\nu_e$ appearance data of T2K \cite{Abe:2011sj}.
  The $\nu_e$ appearance data of T2K prefers $\delta_{CP}$ to be in the lower half plane. From eq.~(\ref{eqexact}), this indicates that $\zeta$ should be positive. The best fit value of this data is equal to -$\pi/2$ which prefers that perturbation in the charged lepton sector is extremely small. The value of $\delta_{CP}$ depends on the relative strengths of the perturbations, $s$ in the charged lepton sector and $\zeta$ in the neutrino sector. This dependence is plotted in figure~\ref{neutrCP}. From this 
  figure, we note that $\zeta \geq 2 s$ if $\delta_{CP} \geq \pi/4$ and $\delta_{CP}$ quickly becomes very small for $\zeta < s$. Figure~\ref{neutrangles} shows the variation of mixing angles with respect to $\zeta$ where the bands for $1\sigma$ and $2\sigma$ bounds are also drawn. The value of $\sin^2\theta_{13} \approx 0.025$ near $\zeta \approx 0.36$. For this value of $\zeta$, the change in $\sin^2\theta_{12}$ is negligibly small($\sim 3\%$). Eventhough the value of $\zeta$ is moderately large, the parameter $a= \zeta M'/M$ quantifying the perturbation in the neutrino sector is quite small because $M'\ll M$. The value of $\sin^2\theta_{23}$ remains $0.5$ if the perturbation in the charged lepton sector is zero. 
  
  In the present scenario, there is a tension between obtaining a large $\delta_{CP}$ and a 
  value of $\sin^2 \theta_{23}$ close to the best fit experimental value. This value of $\sin^2\theta_{23}$ is $15\%$ larger than TBM value of 0.5. In order to obtain this large a deviation, we need a value of $s \approx 0.19$ in the charged lepton sector. For $s \approx 0.19$, the constraint on $\sin^2 \theta_{13}$ in eq.~(\ref{th13}) leads to very small values of $\zeta$ and hence of $\delta_{CP}$. A large CP violation, on the other hand, requires $\zeta > 2 s$, which keeps the value of $\sin^2 \theta_{23}$ 
  close to the TBM value of $0.5$, as can be seen from eq.~(\ref{th23}). Current experiments T2K and
  NO$\nu$A can improve the precision on $\sin^2 \theta_{23}$. If the central value comes closer to 
  $0.5$, then it is possible to have large CP violation. Otherwise, the CP violation is constrained to 
  remain small in this scenario.
 %The amount of the perturbation is related to the value of $M'$ through the equation $a/\zeta=M'/M$. 
\newpage
\section{Summary and Conclusion}

We consider the phenomenology of a model with $A_4$ symmetry which predicts the
tribimaximal form for the PMNS matrix. In this model, we have
introduced $Z_2 \times Z_2$ invariant perturbations in both
the charged lepton and the neutrino sectors. We find that
perturbations in the charged lepton sector alone ($\zeta=0$) can lead to
acceptable values of $\theta_{13}$ but do not give any
CP violation. But, perturbations purely in the neutrino
sector ($s=0$) give rise to viable values of $\theta_{13}$ and
maximal CP violation. Any desired value of the CP violating
phase $\delta_{CP}$ can be obtained by choosing the
appropriate values for the perturbations in the charged lepton
and neutrino sectors. However, there is a tension between the 
requirement to obtain a large CP violation and the need to have the
value of $\sin^2 \theta_{23}$ close to its best fit value. The current 
experiments may be able to settle this issue.
  The final Lagrangian has no overall residual symmetry even though
the neutrino sector has a residual $Z_2$ symmetry.
It will be interesting to explore whether there could be some consequences due to residual symmetry in neutrino sector.

\section*{Acknowledgement}
We are grateful to Walter Grimus for a critical reading of the manuscript.

\bibliographystyle{ieeetr}
\bibliography{referenceslist}

\end{document}